\documentclass[12pt]{article}
\usepackage{epsfig}
\font\bb=msbm10 at 12pt 

\def\rR{\hbox{\bb R}}

\newcommand{\mb}[1]{ \mbox{\boldmath$#1$} }

\newcommand{\beq}{\begin{eqnarray}}
\newcommand{\eeq}{\end{eqnarray}}
\newcommand{\beqq}{\begin{eqnarray*}}
\newcommand{\eeqq}{\end{eqnarray*}}
\newcommand{\p}{\partial}

\newcommand{\x}{\mbox{\boldmath$x$}}
\newcommand{\y}{\mbox{\boldmath$y$}}

\begin{document}

\date{\today }

\pagestyle{empty}

\begin{center}
{\bf ON RECOVERING THE SHAPE OF A DOMAIN FROM THE TRACE OF THE
HEAT KERNEL}

\vspace{5mm} Z. Schuss\footnote{%
Partially supported by a research grant from the Foundation for Basic
Research administered by the Israel Academy of Science and by a research
grant from the US-Israel Binational Science Foundation.}

\vspace{3mm} Department of Mathematics, Tel-Aviv University\\[0pt]
Ramat-Aviv, Tel-Aviv 69978, Israel

\vspace{5mm}A. Spivak

\vspace{3mm} Department of Sciences, Academic Institute of
Technology,\\[0pt]
POB 305, Holon 58102, Israel

\vspace{3mm}

\vspace{5mm}

{\bf ABSTRACT}
\end{center}

The problem of recovering geometric properties of a domain from
the trace of the heat kernel for an initial-boundary value problem
arises in NMR microscopy and other applications. It is similar to
the problem of ``hearing the shape of a drum'', for which a
Poisson type summation formula relates geometric properties of the
domain to the eigenvalues of the Dirichlet or Neumann problem for
the Laplace equation. It is well known that the area,
circumference, and the number of holes in a planar domain can be
recovered from the short time asymptotics of the solution of the
initial-boundary value problem for the heat equation. It is also
known that the length spectrum of closed billiard ball
trajectories in the domain can be recovered from the eigenvalues
or from the solution of the wave equation. This spectrum can also
be recovered from the heat kernel for a compact manifold without
boundary. We show that for a planar domain with boundary, the
length spectrum can be recovered directly from the short time
expansion of the trace of the heat kernel. The results can be
extended to higher dimensions in a straightforward manner.
\newline

\newpage \pagestyle{plain}

\noindent {\large {\bf 1. Introduction}}\newline
\renewcommand{\theequation}{1.\arabic{equation}}
\setcounter{equation}{0} %
\renewcommand{\thefigure}{1.\arabic{figure}} \setcounter{figure}{0}

The problem of recovering geometric properties of a domain from
NMR measurements arises in oil explorations and in non-invasive
microscopy of cell structure \cite{Callaghan}. In these
measurements the trace of the heat kernel for the initial value
problem with reflecting (Neumann) boundary conditions is measured
directly. The problem is analogous to ``hearing the shape of a
drum'', where the solution of the wave equation in the domain is
measured directly (it is ``heard'').

The problem of recovering geometrical properties of a domain from
the eigenvalues of the Dirichlet or Neumann problem for the
Laplace equation in a domain has attracted much attention in the
literature (see \cite{Kac}-\cite{GM} for some history and early
results; for more recent work see \cite{Berry}, \cite{Zelditch} an
references therein).

The mathematical statement of the problem is as follows. Green's
function for the heat equation in a smooth planar domain $\Omega$,
with homogeneous Dirichlet boundary conditions, satisfies
\begin{eqnarray}
\frac{\p G(\y,\x,t)}{\p t} &=&D\Delta_{\y}G(\y,\x,t)
\quad\mbox{for $\y,\x \in\Omega,\ t>0$}
\label{heq} \\
&&\nonumber\\
 G(\y,\x,0)&=&\delta(\mbox{\boldmath$y
$}-\x)  \label{ic} \\
&&\nonumber\\
 G(\y,\x,t)&=&0\quad\quad\quad\quad \quad
\quad\quad\mbox{for $\y\in\p \Omega,\ \x \in\Omega,\ t>0$}.
\label{bcd}
\end{eqnarray}
The function $G(\x,\x,t)\,d\x$ is the probability of return to
$\x\,d \x$ at time $t$ of a free Brownian particle that starts at
the point $\x$ at time $t=0$ and diffuses in $\Omega$ with
diffusion coefficient $1$, with absorption at the boundary
$\p\Omega$. If it is reflected at $\p\Omega$, rather than
absorbed, the Dirichlet boundary condition (\ref{bcd}) is replaced
with the Neumann condition \cite{book}
 \beq
\frac{\p G(\y,\x,t)}{\p\mb{\nu}(\y)}&=&0\quad\quad\quad\quad \quad
\quad\quad\mbox{for $\y\in\p \Omega,\ \x \in\Omega,\ t>0$},
\label{bcn}
 \eeq
where $\mb{\nu}(\y)$ is the unit outer normal at the boundary
point $\y$. The trace of the heat kernel is defined as
\begin{eqnarray}
P(t)=\int_\Omega G(\x,\x,t)\,d%
\x \label{Pt}
\end{eqnarray}
and can be represented by the Dirichlet series
\begin{eqnarray}
P(t)=\sum_{n=1}^\infty e^{-\lambda_nt},  \label{DS}
\end{eqnarray}
where $\lambda_n$ are the eigenvalues of Laplace equation with the
Dirichlet or Neumann boundary conditions (\ref{bcd}) or
(\ref{bcn}), respectively.

It has been shown by Kac \cite{Kac} that for a domain $\Omega $
with smooth boundary $\p \Omega $, the leading terms in the
expansion of $P(t)$ in powers of $\sqrt{t}$ are
\[
P_{\mbox{Kac}}(t)\sim \frac{|\Omega |}{4\pi t}-\frac{|\p \Omega
|}{8%
\sqrt{\pi t}}+\frac{1}{6}(1-r)+O\left( \sqrt{t}\right) ,\quad
\mbox{for
$t\to 0$},
\]%
where $|\Omega |$ denotes the area of $\Omega $, $|\p \Omega |$
denotes the arc-length of $\p \Omega $, and $r$ is the number of
holes in $\Omega $. The full short time asymptotic power series
expansion of $P(t)$ in the form
\[P(t)\sim \sum_{n=0}^{\infty }a_{n}t^{n/2-1},\]
can be deduced from the large $s$ expansion of the Laplace
transform
\[g(s)=\int_{0}^{\infty }\exp \{-s^{2}t\}\left(
P(t)-\frac{a_{0}}{t}\right)\,dt,\quad \left( a_{0}=\frac{|\Omega
|}{4\pi }\right).\]
 in inverse powers of $ s$. Such an expansion
was given by Stewartson and Waechter \cite{SW} in the form
\[\hat{g}(s)\sim \sum_{n=1}^{\infty }\frac{c_{n}}{s^{n}},\]
where
\[c_{n}=a_{n}\Gamma \left( \frac{n}{2}\right) .\]
The constants $c_{n}$ are computable functionals of the curvature of the
boundary. The full expansion is denoted
\begin{equation}
P_{\mbox{SW}}(t)\sim \frac{|\Omega |}{4\pi t}-\frac{|\p \Omega
|}{8 \sqrt{\pi t}}+\frac{1}{6}(1-r)+\sum_{n=3}^{\infty
}a_{n}t^{n/2-1},\quad \mbox{for $t\to 0$}.  \label{sum}
\end{equation}
If the boundary is not smooth, but has cusps and corners, the
expansion contains a term of the order $t^{-\nu }$, where $\nu $
is a number between 0 and 1/2.

The Stewartson-Waechter expansion was used in \cite{Berry} to
deduce further geometric properties of $\Omega$ by extending
$g(s)$ into the complex plane. Examples were given in \cite{Berry}
of the resurgence of the length spectrum of closed billiard ball
trajectories in the domain.

The full length spectrum of closed geodesics on a compact
Riemannian manifold without boundary $\Omega$ appeared in the
short time asymptotic expansion given in \cite{CdV},
 \begin{eqnarray}
P(t)\sim  \frac{1}{\sqrt{\pi t}}\sum_{n=0}^\infty
 P_n(\sqrt{t})\, e^{-\delta_n^2/t}, \quad
\mbox{for $t\to 0$},  \label{cdv}
\end{eqnarray}
where $\delta_n$ are the lengths of closed geodesics on $\Omega$
and $ P_n(x)$ are power series in $x$.

In this paper, we construct an expansion of the form (\ref{cdv})
for the trace of the heat kernel for the initial-boundary value
problem (\ref{heq})-(\ref{bcd}) or (\ref{bcn}) in a smooth bounded
domain $\Omega$ in $\rR^2$. The results can be generalized to
higher dimensions in a straightforward manner.

The point of departure for our analysis is the observation that
transcendentally small terms are not included in the expansion
(\ref{sum}).
These terms have been neglected in \cite{SW} and \cite{Berry} even in
the
case of a circular domain, where the Laplace transform of $G(%
\y,\x,t)$ can be expressed explicitly in terms of modified Bessel
functions. In \cite{Berry} this Laplace transform is expanded in
inverse powers of $s$ and the coefficients $c_n$ are evaluated
asymptotically for large $n$.

A generalization of the asymptotic expansion ``beyond all orders''
(\ref{cdv}) has the form
\begin{eqnarray}
P(t)\sim P_{\mbox{SW}}(t)+ \frac{1}{\sqrt{\pi t}}\sum_{n=1}^\infty%
 P_n(\sqrt{t})\, e^{-\delta_n^2/t}, \quad
\mbox{for $t\to 0$},  \label{bao}
\end{eqnarray}
where $\delta_n$, ordered by magnitude, are constants to be
determined, and $ P_n(x)$ are power series in $x$.
Transcendentally small terms may be, in fact, quite large and make
a finite contribution to the expansion (\ref{bao}) \cite{Meyer}.

To recover the geometrical information from the expansion
(\ref{bao}), given the (measured) function $P(t)$, we note that
\begin{eqnarray}
|\Omega|=\lim_{t\to0}4\pi tP(t),\quad|\p
\Omega|=-\lim_{t\to0}8\sqrt{%
\pi t}\left[ P(t)-\frac{|\Omega|}{4\pi t}\right],
\end{eqnarray}
and so on. This way the entire expansion (\ref{sum}) can be determined.

Once the coefficients of the expansion (\ref{sum}) have been
determined, the exponent of the dominant term of the
transcendentally small part, $\delta_1$, is found as
\[\delta_1=-\lim_{t\to0}\,t\,\log\left[P(t)-P_{\mbox{SW}}(t)\right].\]
Proceeding this way, we can recover the entire expansion
(\ref{bao}) if $P(t) $ is known (e.g., from measurements).

In this paper, we use the ``ray method'', as developed in
\cite{CL}-\cite{Charlie}, to construct a short time asymptotic
expansion of the heat kernel. We use it to expand the trace
asymptotically beyond all orders (the so called ``hyperasymptotic
expansion") and show that the exponents $\delta_i$ are the squares
of half the lengths of the periodic orbits in the domain. The
exponentially small terms in the expansion (\ref{bao}) are due to
rays reflected in the boundary, much like in the geometric theory
of diffraction \cite{Keller1}-\cite{Keller3}. This recovers the
length spectrum of closed billiard ball trajectories in the
domain. In particular, the smallest exponent $\delta_1$ is the
width of the narrowest bottleneck in the domain.\newline

\noindent {\large {\bf 2. The one-dimensional case}}\newline
\renewcommand{\theequation}{2.\arabic{equation}}
\setcounter{equation}{0} %
\renewcommand{\thefigure}{2.\arabic{figure}} \setcounter{figure}{0}

The solution of the heat equation in an interval can be constructed by
the
method of images. Specifically, the Green function of the problem
satisfies
\begin{eqnarray}
\frac{\p G(y,x,t)}{\p t}&=&\frac{\p^2 G(y,x,t)}{\p
y^2}\quad\mbox{for $0<x,y<a,\ t>0$}  \label{1dheq} \\
&&  \nonumber \\
G(y,x,0)&=&\delta(y-x)\quad\quad\mbox{for $0<x,y<a$}  \label{1dic} \\
&&  \nonumber \\
\left(\frac{\p}{\p y}\right)^k
G(0,x,t)&=&\left(\frac{\p}{%
\p y}\right)^kG(a,x,t)=0\quad\mbox{for $0<x<a,\ t>0,\ k=0,1$}.
\label{1dbc}
\end{eqnarray}
The method of images gives the representation
\begin{eqnarray}
&&G(y,x,t) =\label{1dsum} \\
&&\frac{1}{2\sqrt{\pi t}}
\sum_{n=-\infty}^\infty\left[\exp\left\{-\frac{%
(y-x+2na)^2}{4t}\right\}
-(-1)^k\exp\left\{-\frac{(y+x+2na)^2}{4t}\right\}%
\right],\quad(k=0,1).  \nonumber
\end{eqnarray}
Note that if the infinite series is truncated after a finite number of
terms, the boundary conditions are satisfied only in an asymptotic sense
as $%
t\to0$. That is, the boundary values of the truncated solution
decay exponentially fast in $t^{-1}$ as $t\to0$ and the
exponential rate increases together with the number of retained
terms.

The trace is given by
\begin{eqnarray}
\int_0^aG(x,x,t)\,dx&=&\frac{1}{2\sqrt{\pi t}}
\int_0^a\sum_{n=-\infty}^%
\infty\left[\exp\left\{-\frac{(na)^2}{t}\right\}
+(-1)^k\exp\left\{-\frac{%
(x+na)^2}{t}\right\}\right]\,dx  \nonumber \\
&=&\frac{1}{2\sqrt{\pi t}}
\sum_{n=-\infty}^\infty\left[a\exp\left\{-\frac{%
(na)^2}{t}\right\}
+(-1)^k\int_0^a\exp\left\{-\frac{(x+na)^2}{t}\right\}\,dx%
\right]  \nonumber \\
&=&\frac{a}{2\sqrt{\pi t}}\sum_{n=-\infty}^\infty
\exp\left\{-\frac{(na)^2}{t%
}\right\}+\frac{(-1)^k}2  \nonumber \\
&=&\frac{a}{2\sqrt{\pi t}}+\frac{(-1)^k}2+\frac{a}{2\sqrt{\pi t}}%
\sum_{n\neq0}\exp\left\{-\frac{(na)^2}{t}\right\},\quad(k=0,1).
\label{1dprto}
\end{eqnarray}
On the other hand,
\begin{eqnarray}
\int_0^aG(x,x,t)\,dx=\sum_{n=1}^\infty e^{-\lambda_nt},
\label{1dsumlambda}
\end{eqnarray}
where $\{\lambda_n\}$ are the eigenvalues of the homogeneous Dirichlet
or
Neumann problem for the operator $d^2/dx^2$ in the interval $[0,a]$.
Thus
\begin{eqnarray}
\sum_{n=1}^\infty e^{-\lambda_nt}=\frac{a}{2\sqrt{\pi t}}
+\frac{(-1)^k}2+%
\frac{a}{2\sqrt{\pi t}}\sum_{n\neq0} \exp\left\{-\frac{(na)^2}{t}%
\right\},\quad(k=0,1).  \label{1dexpansion}
\end{eqnarray}

If instead of a single interval of length $a$, we consider the
heat equation in a set $\Omega$ consisting of $K$ disjoint
intervals of lengths $l_j,\ (j=1,..,K)$, respectively, the
resulting expansion is
\begin{eqnarray}
\sum_{n=1}^\infty
e^{-\lambda_nt}=\frac{\sum_{j=1}^Kl_j}{2\sqrt{\pi t}}
+(-1)^k\frac{2K}{4}+\sum_{j=1}^K\frac{l_j}{2\sqrt{\pi t}}
\sum_{n\neq0}\exp%
\left\{-\frac{(nl_j)^2}{t}\right\},\quad(k=0,1).  \label{1dD}
\end{eqnarray}
The numerator in the first term on the right hand side of eq.(\ref{1dD})
can
be interpreted as the ``area'' of $\Omega$, so we denote it $%
\sum_{j=1}^Kl_j=|\Omega|$. The number $2K$ is the number of
boundary points of $\Omega$, which can be interpreted as the
``circumference" of the boundary, so we denote it $|\p
\Omega|=2K$. The exponents in the sum on the right hand side of
eq.(\ref{1dD}) can be interpreted as the ``widths" of the
components of $\Omega$. Clearly, for small $t$, the term
containing the smallest width, $r=\min_{1\leq j\leq K}l_j$, will
dominate the sum. Thus we can rewrite eq.(\ref{1dD}) as
\begin{eqnarray}
\sum_{n=1}^\infty e^{-\lambda_nt}=\frac{|\Omega|}{2\sqrt{\pi t}}
-\frac{%
|\p \Omega|}{4}+\frac{mr}{\sqrt{\pi t}} \exp\left\{-\frac{r^2}{t}%
\right\}+ \sum_{l_j> r}\frac{l_j}{2\sqrt{\pi t}}
\sum_{n\neq0}\exp\left\{-%
\frac{(nl_j)^2}{t}\right\},  \label{1dDg}
\end{eqnarray}
where $m$ is the number of the shortest intervals in $\Omega$.

Equation (\ref{1dDg}) can be viewed as the short time asymptotic
expansion of the sum on the left hand side of the equation. The
algebraic part of the expansion consists of the first two terms
and all other terms are transcendentally small. The geometric
information in the various terms of the expansion consists of the
``area'' of $\Omega$ and the ``circumference'' $|\p \Omega|$ in
the algebraic part of the expansion. The transcendental part of
the expansion is dominated by the term containing the smallest
``width'' of the domain, $r$.

The geometric information about $\Omega$ contained in the
algebraic part is the information given in the ``Can one hear the
shape of a drum'' expansions \cite{Kac}, \cite{SW}. The geometric
information contained in the transcendentally small terms in
(\ref{1dDg}) can be understood as follows. The terms $nl_j$ in the
exponents are the lengths of closed trajectories of billiard balls
in $\Omega$, or the lengths of closed rays reflected at the
boundaries, as in \cite{BalianBloch}.

The representation (\ref{1dsum}) can be constructed as a short
time approximation to the solution of the heat equation
(\ref{1dheq})-(\ref{1dbc}) by the {\em ray method} \cite{CL}. In
this method the solution is constructed in the form
\begin{eqnarray}
G(y,x,t)=e^{-\displaystyle S^2(y,x)/4t}\sum_{n=0}^\infty
Z_n(y,x)t^{n-1/2}.  \label{1dray}
\end{eqnarray}
Substituting the expansion (\ref{1dray}) into the heat equation
(\ref{1dheq}%
) and ordering terms by orders of magnitude for small $t$, we obtain at
the
leading order the {\em ray equation}, also called the {\em eikonal
equation}%
,
\begin{eqnarray}
\left|\frac{\p S(y,x)}{\p y}\right|^2=1, \label{1deikonal}
\end{eqnarray}
and at the next orders the {\em transport equations}
\begin{eqnarray}
&&2\frac{\p S(y,x)}{\p y}\frac{\p Z_n(y,x)}{\p y}
+Z_n(y,x)\left(\frac{\p^2 S(y,x)}{\p y^2}+\frac{2n}{S(y,x)}%
\right)  \nonumber \\
&&\frac{2}{S(y,x)}\frac{\p^2Z_{n-1}(y,x)}{\p y^2},\quad
n=0,1,\dots\,.  \label{1dtransport}
\end{eqnarray}
Denoting
\[
p(y,x)=\frac{\p S(y,x)}{\p y},
\]
we write the equations of the {\em characteristics}, or {\em rays} of
the
eikonal equation (\ref{1deikonal}) as \cite{CourantHilbert}
\begin{eqnarray}
\frac{\p y(\tau,x)}{\p t}=2p,\quad \frac{d
p(\tau)}{dt}=0,\quad%
\frac{dS(\tau)}{d\tau}=2p^2(\tau)  \label{1drays}
\end{eqnarray}
with the initial conditions
\begin{eqnarray*}
y(0,x)=x,\quad p(0)=\pm1,\quad S(0)=0.
\end{eqnarray*}
The condition $S(0)=0$ is implied by the initial condition $%
G(x,y,0)=\delta(x-y)$. The solutions are given by
\begin{eqnarray}
y(\tau,x)=x+2p\tau,\quad p(\tau)=\pm1,\quad
S(\tau)=2\tau=\pm(y-x).
\end{eqnarray}
Thus $S(y,x)$ is the length of the ray from $y$ to $x$. We denote this
solution by $S_0(y,x)$. It is easy to see that the solution of the
transport
equations corresponding to $S_0(y,x)$ is given by $Z_0(y,x)=const$,
and $%
Z_n(y,x)=0$ for all $n\geq1$. The initial condition (\ref{1dic})
implies
that
\[
Z_0(y,x)=\frac{1}{2\sqrt{\pi}}.
\]
Combined in eq.(\ref{1dray}) this solution gives Green's function for
the
heat equation on the entire line,
\[
G_0(y,x,t)=\frac{1}{2\sqrt{\pi
t}}\exp\left\{-\frac{(y-x)^2}{4t}\right\},
\]
which is the positive term corresponding to $n=0$ in the expansion
(\ref%
{1dsum}).

The ray from $x$ to $y$ is not the only one emanating from $x$. There
are
rays emanating from $x$ that end at $y$ after reflection in the
boundary.
Thus the ray from $x$ that reaches $y$ after it is reflected at the
boundary
$0$ has length $y+x$. Therefore there is another solution of the eikonal
equation, $S_1(y,x)$, which is the length of the reflected ray, given by

\[
S_1(y,x)=y+x.
\]
The ray from $x$ that reaches $y$ after it is reflected at the boundary
$a$
has length $2a-x-y$. The ray from $x$ to $0$, then to $a$, and then to
$y$
has length $2a+x-y$. Thus the lengths of all rays that reach $y$ from
$x$
after any number of reflections in the boundary generate solutions of
the
eikonal equation, which are the lengths of the rays, which in turn
generate
solutions of the heat equation. We denote them by $S_k(y,x)$ with some
ordering. The corresponding solutions of the transport equation are
\[
Z_{0,k}(y,x)=\frac{C_k}{2\sqrt{\pi}},
\]
where $C_k$ are constant. They are chosen so that the sum of all the ray
solutions,
\[
G_k(y,x,t)=\frac{Z_{0,k}(y,x)}{\sqrt{t}}e^{-\displaystyle
S_k^2(y,x)/4t},
\]
satisfies the boundary conditions (\ref{1dbc}). Note that for all
$k\neq0$
\[
G_k(y,x,t)\to0\quad\mbox{as $t\to0$}.
\]
This construction recovers the solution (\ref{1dsum}).\newline

\noindent {\large {\bf 3. The ray method for short time asymptotics of
Green's function}}\newline
\renewcommand{\theequation}{3.\arabic{equation}}
\setcounter{equation}{0} %
\renewcommand{\thefigure}{3.\arabic{figure}} \setcounter{figure}{0}

The ray method consists in the construction of Green's function $G(%
\y,\x,t)$ in the asymptotic form
\begin{eqnarray}
G(\y,\x,t)\sim e^{-\displaystyle
S^2(\y,%
\x)/4t}\sum_{n=0}^\infty Z_n(\y,%
\x)t^{n-1}.  \label{Raysolution}
\end{eqnarray}
The function $S(\y,\x)$ is the solution of the {\em eikonal
equation}
\begin{eqnarray}
\left|\nabla_{\y}
S(\y,\x%
)\right|^2=1  \label{eikonaleq}
\end{eqnarray}
and the functions $Z_n(\y,\x)$ solve the {\em transport equations}
\begin{eqnarray}
&&2\nabla_{\y}S(\y, \x)\cdot \nabla_{\y}
Z_n(\y,%
\x)+ Z_n(\y,\x)\left[ \Delta_{\y}
S(\y,\x)+%
\frac{2n-1} {S(\y,\x)}\right]= \nonumber
\\
&&\frac{2} {S(\y,\x)} \Delta_{%
\y}Z_{n-1}(\y,\x),
\quad%
\mbox{for $n=0,1,2,\dots$.}  \label{transport}
\end{eqnarray}

The eikonal equation (\ref{eikonaleq}) is solved by the {\em method of
characteristics} \cite{CourantHilbert}. The characteristics, called {\em
rays%
}, satisfy the differential equations
\begin{eqnarray}
\frac{d\y(\tau,\x)}{d\tau}=
2\nabla_{%
\y} S(\y(\tau,\x),%
\x), \quad\frac{d\nabla_{\y} S(%
\y(\tau,\x),
\x)}{d\tau}%
=0,\quad \frac{dS(\y(\tau,\x),%
\x)}{d\tau}= 2.  \label{characteq}
\end{eqnarray}
The initial condition (\ref{ic}) implies that the rays emanate
from the point $\x$. Thus we choose the initial conditions
\begin{eqnarray}
\y(0,\x)=\x,\quad
\nabla_{%
\y} S(\y(0,\x),%
\x)= \mbox{\boldmath$\nu$},\quad
S(\y(0,%
\x),\x)=0,  \label{cic}
\end{eqnarray}
where $\mbox{\boldmath$\nu$}$ is a constant vector of unit length. The
solution is given by
\begin{eqnarray}
\y(\tau,\x)= \x+2%
\mbox{\boldmath$\nu$}\tau,\quad
S(\y,\x)=|%
\y- \x|= 2\tau,\quad \nabla_{%
\y} S(\y,\x)= %
\mbox{\boldmath$\nu$}.  \label{characteristics}
\end{eqnarray}
The pair $(\tau,\mbox{\boldmath$\nu$})$ determines uniquely the point $%
\y=\y(\tau,\x)$ and the value of $S(\y,\x)$ at the point. The
parameter $\tau$ is half the distance from $\y$ to $%
\x$ or half the length of the ray from $\x$ to $\y$. The vector
$\mbox{\boldmath$\nu$}$ is the unit vector in the direction from
$\x$ to $\y$.

The function $Z_0(\y,\x)$ is easily seen to be a constant,
$1/4\pi$, and
$Z_n(\y,\x%
)=0$ for all $n>0$. This construction recovers the solution of the
heat equation in the entire plane and disregards the boundary $\p
\Omega$, because in the plane every point can be seen from every
other point by a straight ray. Note that to calculate the function
$P(t)$ in eq.(\ref{Pt}) only the values of $S(\x,\x)$ and
$Z_0(%
\x,\x)$ are needed. Thus $S(%
\x,\x)=0$ and the first
approximation to $%
G(\x,\x,t)$ is
\begin{eqnarray*}
G(\x,\x,t)=\frac{1}{4\pi t},
\end{eqnarray*}
hence the first approximation to $P(t)$ is
\begin{eqnarray*}
P_0(t)=\frac{|\Omega|}{4\pi t}.
\end{eqnarray*}

There is another solution of the eikonal equation
(\ref{eikonaleq}) constructed along rays that emanate from $\x$,
but
reach $%
\y$ after they are reflected in $\p \Omega$ \cite{CL}. The law of
reflection is determined from the boundary conditions. Dirichlet
and Neumann boundary conditions imply that the angle of incidence
equals that of reflection \cite{CL}. Similarly, there are
solutions of the eikonal equation that are the
lengths of rays that emanate from $\x$ and reach $%
\y$ after any number of reflections in $\p \Omega$. We denote
these solutions $S_k(\y,\x)$ with some ordering. Thus the full ray
expansion of Green's function has the form
\begin{eqnarray}
G(\y,\x,t)\sim \sum_{k=1}^{\infty}e^{-\displaystyle S^2_{k}(\y,
\x)/4t}Z_{k}(\y,\x,t), \label{Raysolution1}
\end{eqnarray}
where
\[Z_{k}(\y,\x,t)=
\sum_{n=0}^{\infty }Z_{n,k} (\y,\x)t^{n-1}.  \] As above, each one
of the series
\[e^{-\displaystyle S_{k}^{2}(\y,
\x)/4t}Z_{k}(\y,\x,t)
\]
is called a {\em ray solution} of the diffusion equation. The
boundary values of $Z_{k}(\y,\x,t)$ are chosen so that
$G(\y,\x,t)$ in eq.(\ref{Raysolution1} ) satisfies the imposed
boundary condition.
In particular, the values of $%
S_{k}(\x,\x)$ are the lengths of all rays that emanate from $\x$
and are reflected from the boundary back to $\x$. Note that sums
of ray solutions satisfy boundary conditions only at certain
points.

To fix the ideas, we consider first simply connected domains. We denote
\[
S_{0}(\y,\x)=\left|
\x-%
\y\right|
\]%
and%
\[
G_{0}(\y,\x,t)=\frac{1}{4\pi t}%
e^{-\displaystyle S_{0}^{2}(\y,\x)/4t}.
\]
We consider first solutions corresponding to
rays that are reflected only once at the boundary, and in particular,
rays that are reflected back from the boundary to the points of their
origin. Such rays hit the boundary at right angles (see Fig. 1 and \cite{CL}).
If there is only one minimal eikonal $S_{1}(\x,%
\x)>0$, we say that $\x$ is a {\em regular}
point of $\Omega $. If there is more than one minimal eikonal $S_{1}(%
\x,\x)$, we say that $\x$ is a {\em critical} point of $\Omega $.
We denote by $\Gamma $ the locus of
critical points in $\Omega $. The eikonal $S_{1}(\y,%
\x)$ is the length of the shortest ray from $\x$ to $\y$ with one
reflection in the boundary such that the ray from $\x$ to the
boundary does not intersect $\Gamma $. For
$\x=\y$ the eikonal $S_{1}(%
\x,\x)$ is twice the distance of $%
\x$ to the boundary. We denote by $\x'$ the orthogonal projection
of $\x$ on the boundary along the shortest normal from $\x$ to the
boundary. When $\y=\x'$
\begin{equation}
S_{1}(\x',\x)=S_{0}(%
\x',\x)=\left| \x-%
\x'\right| .  \label{x-y}
\end{equation}%
The function
\[
G_{1}(\y,\x,t)=e^{-\displaystyle
S_{1}^{2}(%
\y,\x)/4t}Z_{1}\left(
\y,%
\x,t\right)
\]
has to be chosen so that $G_{0}(\x',\x,t)- G_{1}(\x',\x,t)=0$. In
view of (\ref{x-y}), we have to choose
\[Z_{1}\left( \x',\x,t\right)
=\frac{1}{4\pi t}. \] When $\y''$ is the other boundary point on
the normal from $\x'$ to $\x$, we have
\begin{eqnarray}
G_{0}(\y'',\x,t)-\ G_{1}(%
\y'',\x,t) =\frac{1}{4\pi t}%
e^{-\left| \x-\y''%
\right|^{2}/t}- e^{-\left( \left| \x'-\x\right| +\left|
\y''-\x'\right|
\right)^2/t}Z_{1}\left( \y'',%
\x,t\right). \label{error}
\end{eqnarray}

Next, we consider in $\Omega -\Gamma $ the minimal among the
remaining eikonals
$S_{k}(\mbox{\boldmath$x,x$})>S_{1}(\mbox{\boldmath$x,x$})$ and
denote it $S_{2}(\mbox{\boldmath$x,x$})$. This eikonal is twice
the length of a ray that emanates from $\x$, intersects $\Gamma $
once, and intersects the
boundary $\p \Omega $ at right angles at a point, denoted $%
\x''$. The eikonal $S_{2}(\y,\x)$ is the length of the ray from
$\x$ to $\y$
with one reflection in the boundary such that the ray from $%
\x$ to the boundary intersects $\Gamma $ once. When $%
\y=\x''$
\begin{equation}
S_{2}(\x'',\x)=S_{0}(%
\x'',\x)=\left| %
\x-\x''\right| .  \label{T3}
\end{equation}

When $\y^{\prime}$ is the other boundary point on the
normal that emanates from $\x^{\prime\prime}$ (see Fig.2), we have%
\[
S_{2}(\y',\x)=\left| %
\x-\x''\right| +\left| %
\y'-\x''\right|.
\]
In general $\x'\neq\y'$ and $\x''\neq\y''$. However, if the ray is
a 2-periodic orbit (that hits the boundary at only 2 points),
$\x'=\y'$ and $\x''=\y''$ so that
\begin{eqnarray*}
S_{2}(\y'',\x)=S_{0}(%
\y'',\x)=\left|\x- \y''\right|
\end{eqnarray*}
and
\begin{eqnarray*}
S_{2}(\y',\x)= \left|\x-\x''\right| +\left|\y''-\x'\right|.
\end{eqnarray*}
\newpage
\centerline{\epsfig{figure=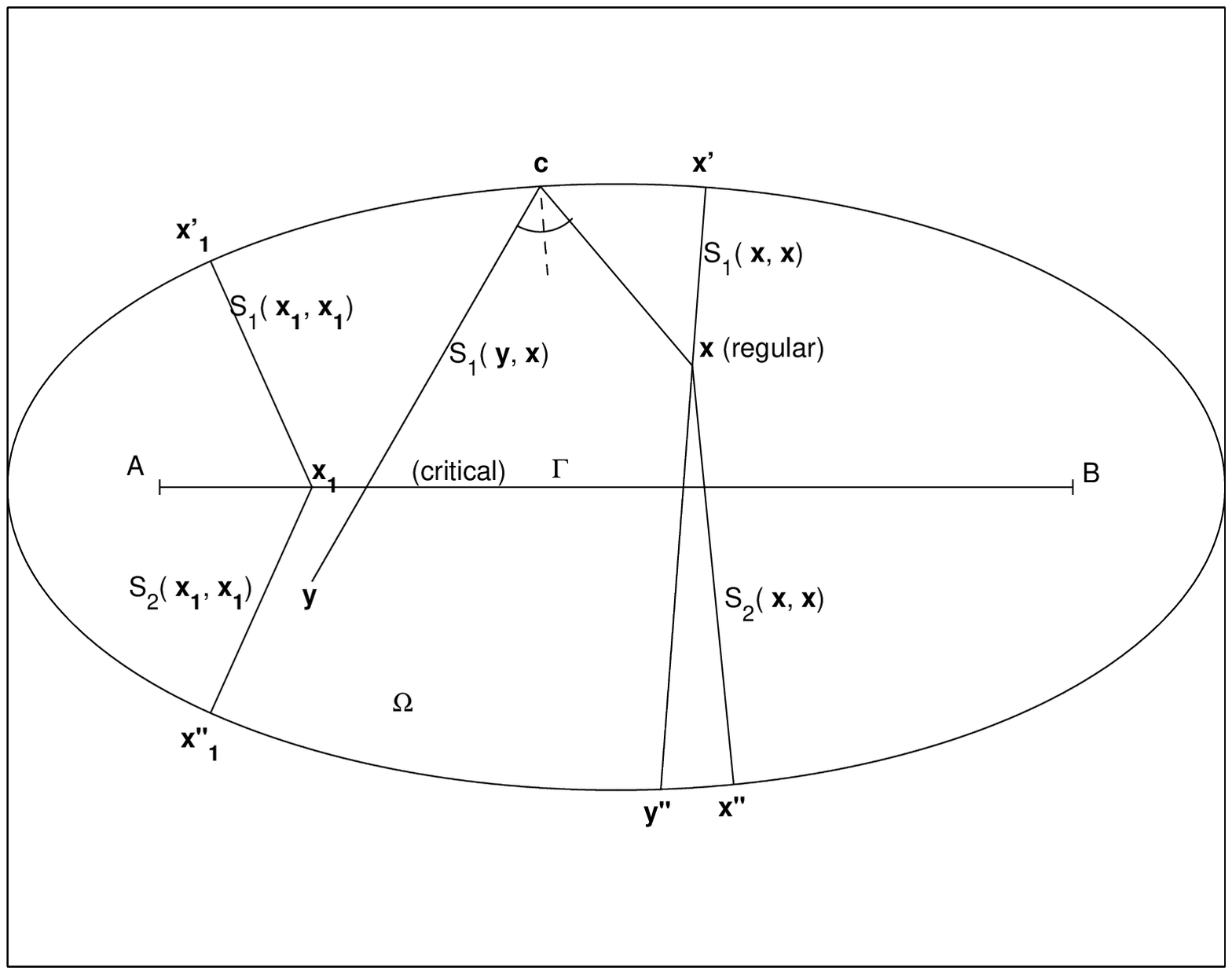,width=5in,height=4in}}
{\it Figure 1. The locus of critical points, $\Gamma$, is  the
segment AB. The first eikonal is
$S_1(\mbox{\boldmath$y,x$})=|\mbox{\boldmath$x-c$}|+|\mbox{\boldmath$c-y$}|$.
It is defined as the shortest reflected ray from $\x$ to $\y$,
such that $\mbox{\boldmath$x-c$}$ does not intersect $\Gamma$. For
$\mbox{\boldmath$x=y$}$ the diagonal values are
 $S_1(\mbox{\boldmath$x,x$})= 2|\mbox{\boldmath$x-x'$}|$. The diagonal values of the
second eikonal are
$S_2(\mbox{\boldmath$x,x$})=2|\mbox{\boldmath$x-x''$}|$. The
vectors $\mbox{\boldmath$x-x'$}$ and $\mbox{\boldmath$x-x''$}$
are orthogonal to the boundary. For ${\x}_1\in\Gamma$ the two
eikonals are equal. }
 \vspace{3mm}

\noindent Since
\[\left|\x-\y''\right|<
\left|\x-\x'\right| +\left|\y''-\x'\right|<\left| \x-\x''\right|
+\left|\y''-\x'\right|
\]
for all regular points $\x$, the order of magnitude of the
boundary error (\ref{error}) decreases if we use the approximation
\begin{eqnarray}
G_0(\y,\x,t)\sim G_0(\y,\x,t)- G_1(\y,\x,t)- G_2(\y,\x,t)
\label{g012}
\end{eqnarray}
with
\[
Z_2(\y'',\x,t)= Z_1(\y'',\x,t)=Z_0(t).
\]

\centerline{\epsfig{figure=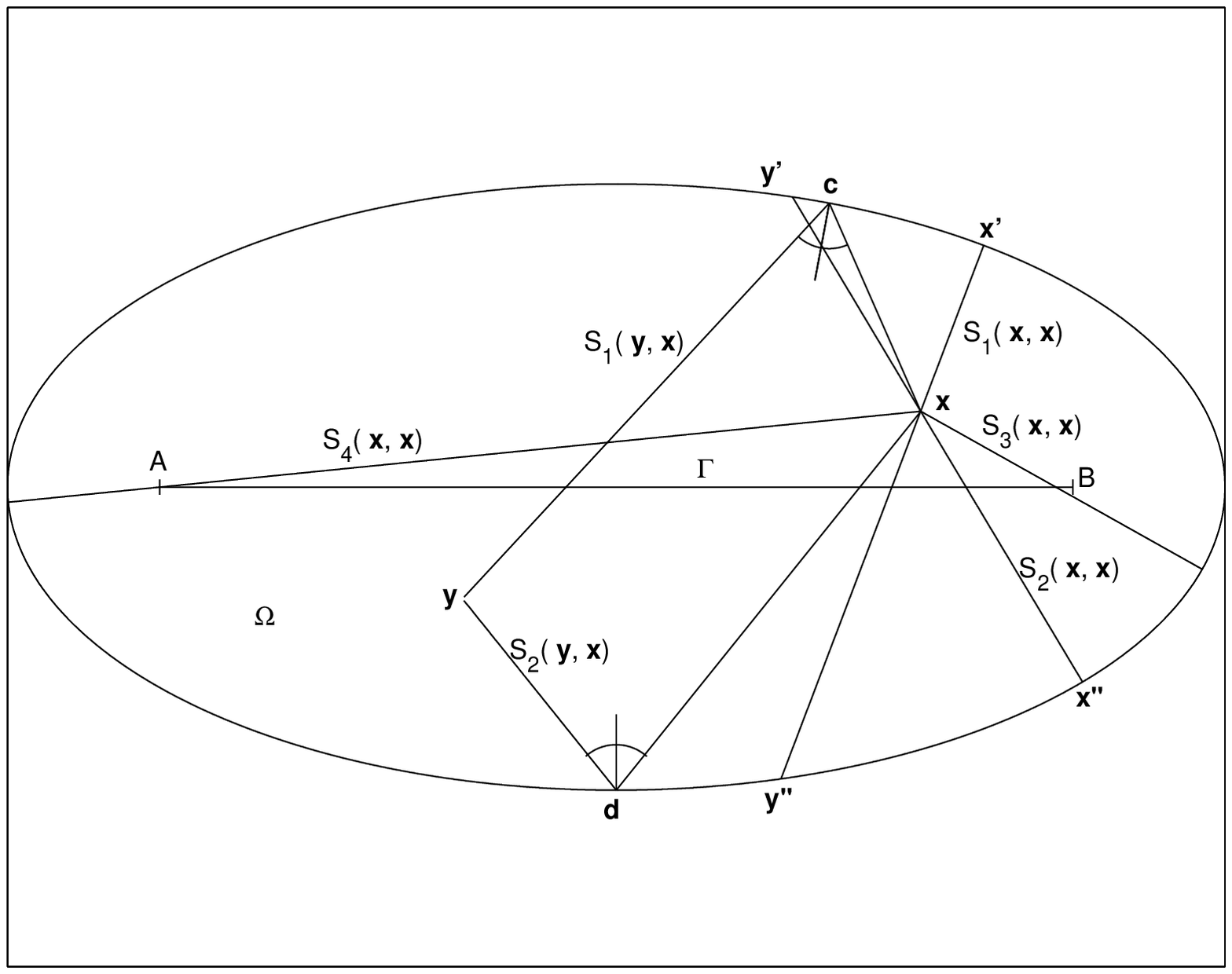,width=5in,height=4in}}
{\it Figure 2. The second eikonal $S_2(\mbox{\boldmath$y,x$})=|\mbox{\boldmath$x-d$}|
+|\mbox{\boldmath$d-y$}|$. It is defined as the shortest reflected ray  such that
$\mbox{\boldmath$x-d$}$ intersects $\Gamma$. The eikonals $S_3(\mbox{\boldmath$x,x$})$
and $S_4(\mbox{\boldmath$x,x$})$ are ordered according to magnitude.}

\vspace{6mm}

\vspace{3mm}

\noindent {\large {\bf 4. The trace}}\newline
\renewcommand{\theequation}{4.\arabic{equation}}
\setcounter{equation}{0} %
\renewcommand{\thefigure}{4.\arabic{figure}} \setcounter{figure}{0}

To find the short time asymptotics of the Dirichlet series (\ref{DS}),
as
given in eq.(\ref{Pt}),
\[
P(t)=\int_\Omega G(\x,\x,t)\,d%
\x,
\]
we use the ray expansion (\ref{Raysolution1}) for the evaluation of the
integral. We retain in the resulting expansion only terms that are
transcendentally small, since all algebraic terms are contained in the
expansion (\ref{sum}).\\

\noindent
{\bf 4.1 Simply connected domains}\\

We note that according to Sard's theorem, $\Gamma $ is a set of
measure zero and that all points in the domain $\Omega -\Gamma $
are regular. For any point $\x\in \Omega $, we denote by
$r_{1}(\x%
)$ its distance to the boundary and note that
$S_{1}(\x,%
\x)=2r_1(\x)$. We also denote by
$s_1(%
\x)$ the arclength at the boundary point $%
\x'$ (the orthogonal projection of $%
\x$ on $\p \Omega $ along the shortest normal
from $%
\x$ to $\p \Omega $), measured from a boundary point where the
arclength is set to 0 (see Figure 3).

\centerline{\epsfig{figure=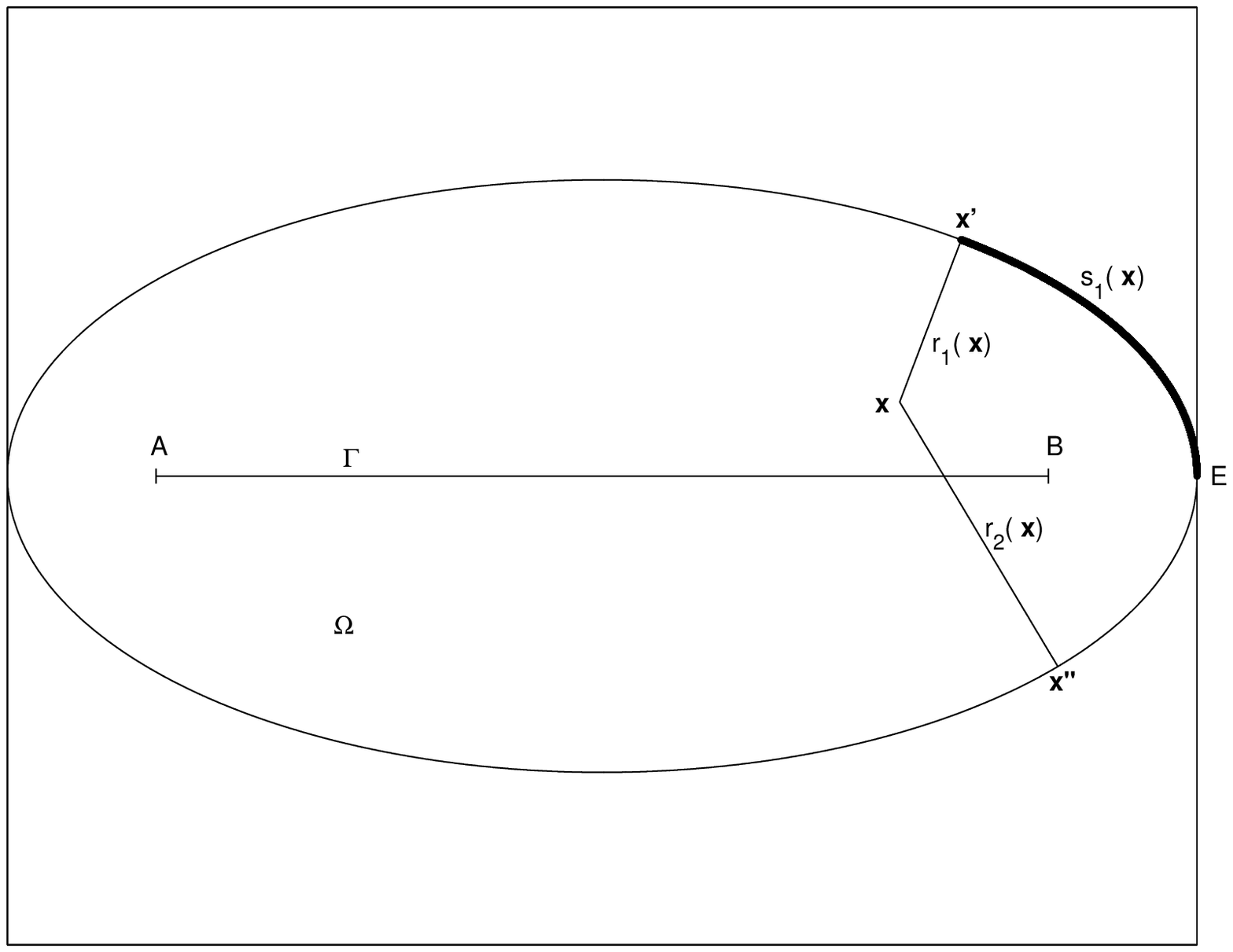,width=5in,height=4in}} {\it
Figure 3. The arclength $s_1(\x)$ is measured from the point $E$.
Both transformations $\x\to(r_1(\x),s_1(\x))$ and
$\x\to(r_2(\x),s_1(\x))$ are one to one mappings of
$\Omega-\Gamma$. The images are given in Figure 4.}

\vspace{6mm}
\noindent
 It follows that the change of variables in
$\Omega
-\Gamma $ , given by
\begin{equation}
\x\rightarrow (r_1(\x),s_1(%
\x)),  \label{T1}
\end{equation}
is a one-to-one mapping of $\Omega -\Gamma $ onto a strip $0\leq
r_1\leq r_1\left( s_1\right),\;0\leq s_1\leq L$, where
$r_1(s_1)$ is the distance from the boundary point corresponding to
arclength $s_1$ to $\Gamma$.

\centerline{\epsfig{figure=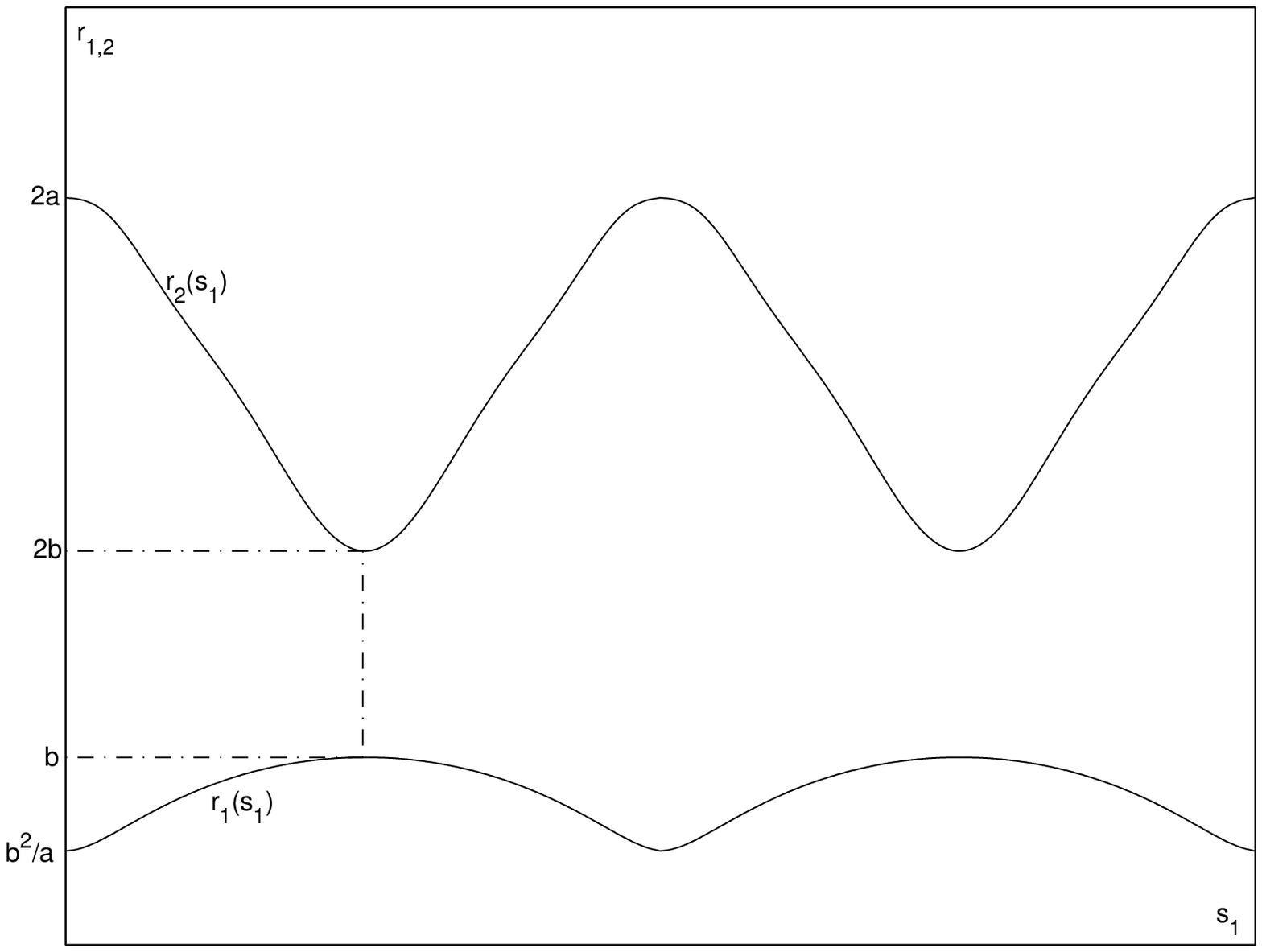,width=5in,height=4in}}
{\it Figure 4. The domain $\Omega$ is the ellipse
$\displaystyle\frac{x^2}{a^2}+\displaystyle\frac{y^2}{b^2}\leq1$. The domain
enclosed between the $s_1$-axis and the lower curve is the image of the ellipse
under the transformation (\ref{T1}) and the domain enclosed between the upper and the lower curves is its image under (\ref{T2}).}

\vspace{6mm}
\noindent
We evaluate the
integral over $\Omega $ separately for each summand $k$ in the expansion
(%
\ref{Raysolution1}). In this notation, we can write
\begin{eqnarray}
&&\int_{\Omega }G_{1}(\x,\x,t)
\,d\x=\nonumber\\
&&\label{tr1}\\
&&\int_{\Omega }e^{-\left[ \displaystyle
S_{1}(\x,\x)%
\right] ^{2}/4t}\sum_{n=0}^{\infty }Z_{n,1}(\x,%
\x)t^{n-1}\,d\x=\int_{0}^{L}\,ds\!%
\int_{0}^{r_1(s_1)}e^{-r_1^{2}/t}J_1(r_1,s_1)Z_{1}(r_1,s_1,t)\,\,dr_1,
\nonumber
\end{eqnarray}
where $J_1(r_1,s_1)$ is the Jacobian of the transformation and
\[
Z_{1}(r_1,s_1,t)=\sum_{n=0}^{\infty
}Z_{n,1}(\mbox{\boldmath$x,x$})t^{n-1}.
\]%
Note that the Jacobian vanishes neither inside $\Omega -\Gamma $ nor at
$r_1=0$,
because the transformation is one-to-one in $\Omega -\Gamma $, however,
it
does on $\Gamma $.

 We set $S_{2}(\mbox{\boldmath$x,x$})=2r_{2}(\x)$ and
use it as a coordinate. We use $s_1(\x)$ as the other coordinate
of the point $\x\in\Omega-\Gamma$. Note that while $r_{2}(\x)$ is
the length of the longer normal from $\x$ to $\p \Omega$ (the one
that intersects $\Gamma$), the other coordinate is the arclength
corresponding to the shorter normal from $\x$ to $\p \Omega$ (the
one that does not intersect $\Gamma$).  The transformation
\begin{equation}
\x\rightarrow \left( r_{2}(\x),s_{1}(\x)\right) \label{T2}
\end{equation}%
maps $\Omega -\Gamma $ onto the strip $r\left( s_{1}\right) \leq
r_{2}\leq
l(s_1) ,\;0\leq s_{1}\leq L $, where $l(s_1)$ is the length of the
segment of the normal that starts at
the boundary point $r_1=0,s_1$  and ends at its other intersection
point with the boundary. This
 mapping is one-to-one as well.
It follows that%
\begin{eqnarray}
&&\int_{\Omega }G_{2}(\x,\x,t)\,d\x=
\nonumber\\
&&\label{tr2}\\
&&\int_{\Omega }e^{-\left[ \displaystyle
S_{2}(\x,\x)%
\right] ^{2}/4t}\sum_{n=0}^{\infty }Z_{n,2}(\x,%
\x)t^{n-1}\,d\x=\int_{0}^{L}\,ds_{1} \!\int_{r(s_{1})}^{l\left(
s_{1}\right)
}e^{-r_{2}^{2}/t}J_{2}(r_{2},s_{1})Z_{2}(r_{2},s_{1},t)\,dr_{2},\nonumber
\end{eqnarray}
where%
\[Z_{2}(r_{2},s_{1},t)=\sum_{n=0}^{\infty
}Z_{n,2}(\mbox{\boldmath$x,x$})t^{n-1}.
\]%
Note that for $\x$ on $\Gamma$ both transformations (\ref{T1}) and
(\ref{T2}) are identical and
\begin{eqnarray*}
J_{2}(r_{2},s_{1})Z_{2}(r_{2},s_{1},t)&=&J_1(r_1,s_1)Z_{1}(r_1,s_1,t).
\end{eqnarray*}
It follows that
 the two equations (\ref{tr1}) and (\ref{tr2}) combine together to give
\begin{eqnarray}
&&\int_{\Omega }\left[G_{1}(\x,\x,t)+G_{2} (\x,\x,t)
\right]\,d\x=\nonumber\\
&&\label{intg1g2}\\
&&\int_0^L\int_0^{l(s)}e^{-r^2/t}J(r,s)Z(r,s,t)\,dr\,ds,\nonumber
\end{eqnarray}
where $s=s_1,\ r=r_1,\ J=J_1,$ and $Z=Z_1$ for $0<r<r_1(s_1)$,
and $s=s_1,\ r=r_2,\ J=J_2$, and $Z=Z_2$ for
$r_2(s_1)<r<l(s_1)$. Thus the domain of integration of the function
$e^{-r^2/t}J(r,s)Z(r,s,t)$ in eq.(\ref{intg1g2}) is the domain enclosed by the $s_1$-axis
and the upper curve in Figure 4. Now, for $t\ll1$, we write the inner integral on the
right hand side of eq.(\ref{intg1g2}) as
\begin{eqnarray*}
&&\int_0^{l(s)}e^{-r^2/t}J(r,s)Z(r,s,t)\,dr=\sqrt{\frac{\pi
t}{2}}\mbox{erf}\left(\frac{l(s)}{\sqrt{t}}\right)J(0,s)Z(0,s,t)\left(1+O
\left(\sqrt{t}\right)\right)=\\
&&\\
&&\sqrt{\frac{\pi
t}{2}}\left(1-\displaystyle\frac{\exp\left\{-\displaystyle\frac{l^2(s)}
{t}\right\}
\sqrt{t}}{l(s)}\right)J(0,s)Z(0,s,t)\left(1+O\left(\sqrt{t}\right)\right)
.
\end{eqnarray*}
Recall that $J(0,s)Z(0,s,t)\neq0$. Only the exponentially small terms
have to be considered, because the
algebraic terms are included in the SW expansion. Thus
\begin{eqnarray*}
&&\int_0^L\int_0^{l(s)}e^{-r^2/t}J(r,s)Z(r,s,t)\,dr\,ds-\int_0^L
\sqrt{\frac{\pi t}{2}}(0,s)Z(0,s,t)
\left(1+O\left(\sqrt{t}\right)\right)\,ds=\\
&&\\
&&-\int_0^L\exp\left\{-\displaystyle\frac{l^2(s)}{t}\right\}\displaystyle
\frac{J(0,s)Z(0,s,t)}{l(s)}
O\left(t\right)\,ds\quad\mbox{for $t\ll1$}.
\end{eqnarray*}
Evaluating the last integral by the Laplace method, we find that each
point $s_i$ that is an extremum point of $l(s)$ contributes and
exponential
term of the form
\begin{eqnarray}
\exp\left\{-\displaystyle\frac{l^2(s_i)}{t}\right\}
\displaystyle\frac{J(0,s_i)Z(0,s_i,t)}{l(s_i)}
O\left(t^{\nu}\right).\label{lsi}
\end{eqnarray}
The expression (\ref{lsi}) means that some of the $\delta_n$-s in the expansion eq.(\ref{bao}) are the extremal
values $l(s_i)$ and their multiples.
These are half the lengths of the 2-periodic orbits of a billiard ball in $\Omega$
(see Figure 5). The 2-periodic orbits of the ellipse are the major axes, which correspond
to the lowest and highest points of the top curve in Figure 4.
There are other exponents as well, as discussed below.

\centerline{\epsfig{figure=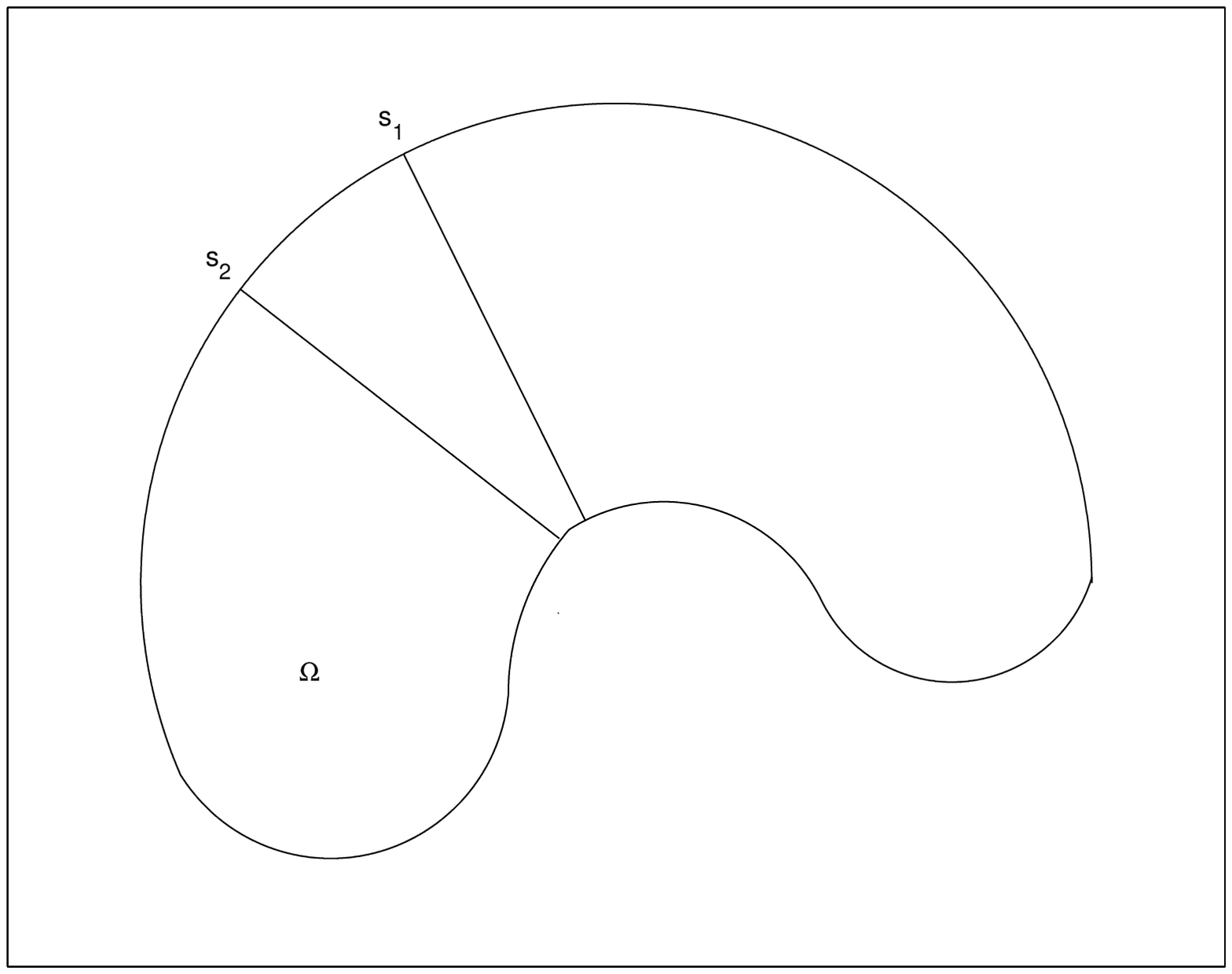,width=5in,height=4in}}
{\it Figure 5. The rays emanating from the boundary points $s_1$ and $s_2$ are orthogonal
 the boundary at both ends. They are 2-periodic orbits.}

\vspace{4mm}
\noindent

The pre-exponential terms in the expression (\ref{lsi}) influence the factors $P_n(\sqrt{t})$ in eq.(\ref{bao}).
For example, if $l'(s_i)=0,\ l''(s_i)\neq0$, then $\nu=3/2$. If the
boundary is flatter, then
$1\leq\nu<3/2$.

In addition to the 2-periodic orbits, there are ray solutions
corresponding to rays from $\x$ to $\y$ that are reflected any
number of times in the boundary. There are eikonals from $\x$ to
$\y$ in $\Omega$ with $N-1$ different vertices on the boundary,
which have $N$ vertices on $\p\Omega$ if $\x=\y$ and
$\x\in\p\Omega$ (this is a periodic orbit with $N-1$ reflections).
Among these periodic orbits there are eikonals
$S_N(\mbox{\boldmath$x,x$})$ with extremal length, denoted
$S_{N,j},\ (j=1,\dots$). At points $\x\in\Omega$ on a 2-periodic
orbit the eikonal $S_N(\mbox{\boldmath$x,x$})$, which now has
$N-1$ vertices on the boundary, may reduce to the 2-periodic orbit
with $N$ reflections. Therefore the change of variables $\x\to
\left(S_N(\x,\x),s(\x) \right)$ will map the domain into a strip
with extremal widths that are the differences between the lengths
$S_{N,j}$ and the length of a 2-periodic orbit with $N$
reflections. It follows that the evaluation of the trace by the
Laplace method leads to exponents which are the extremal lengths
of periodic orbits with any number of reflections.

For example, there is an eikonal in a circle (centered at the
origin) that is the ray from $\x$ to $\y$ with 2 reflections in
the boundary (see Figure 6).

\centerline{\epsfig{figure=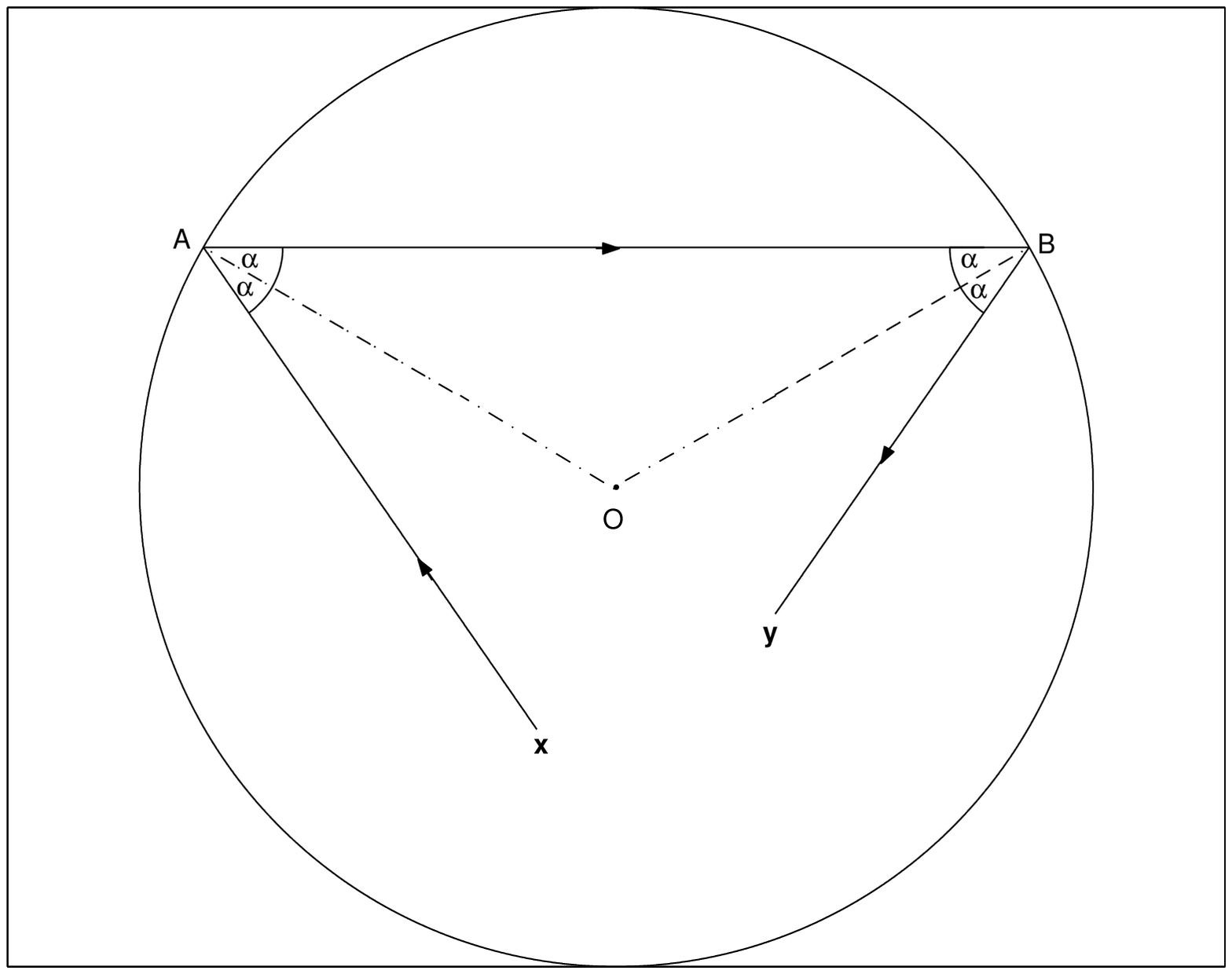,width=5in,height=4in}}
\centerline{\it Figure 6. The eikonal $S_3(\mbox{\boldmath$y,x$})$ with two reflections in the circle.}

\vspace{3mm} \noindent For $\x=\y$ it is the equilateral triangle
(see Figure 7) with circumference
\[S(\x,\x)=R\left(2\frac{\sqrt{2|
\x|^2+1+\sqrt{8|\x|^2+1}}} {\sqrt{4|\x|^2+1+\sqrt{8|\x|^2+1}}}+
{\sqrt{4|\x|^2+2+2\sqrt{8|\x|^2+1}}} \right).\]

\centerline{\epsfig{figure= 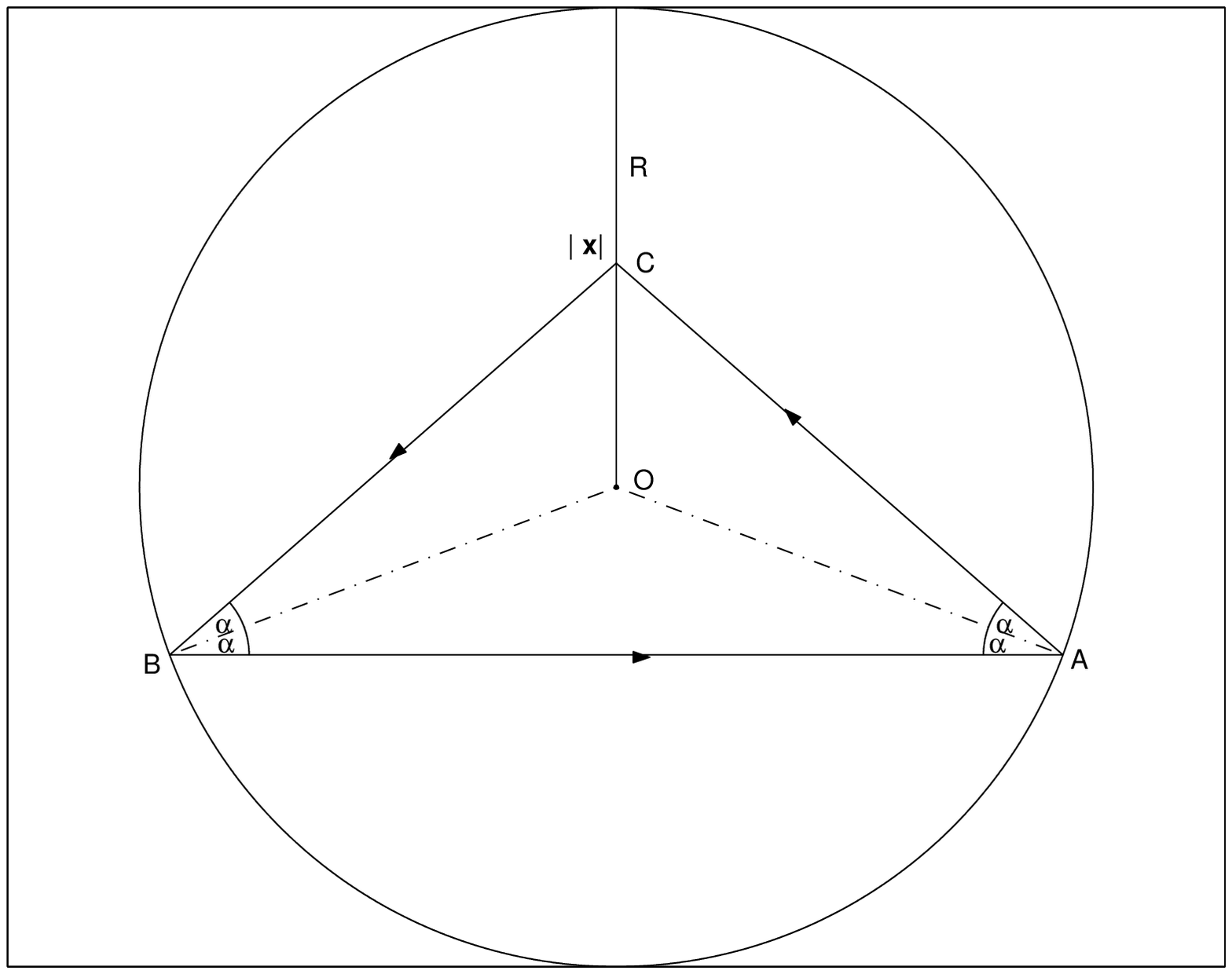,width=5in,height=4in}}
\centerline{\it Figure 7. The eikonal $S_3(\mbox{\boldmath$x,x$})$
with two reflections, where $|\x|=OC$.} \vspace{6mm} \noindent

The eikonal $S_3(\x,\y)$ reduces to a 2-periodic orbit with two
reflections if $\x=\y=0$ (the center of the circle). If $\x$ is on
the circumference, the eikonal becomes the isosceles triangle with
one vertex at $\x$. To evaluate the contribution of the
corresponding ray solution to the trace, we use this eikonal as a
coordinate that varies between $4R$, the length of the 2-periodic
orbit with two reflections, and $3\sqrt{3}R$, the circumference of
the inscribed isosceles triangle. The contribution of this
integral to the exponential sum in eq.(\ref{bao}) contains
exponents that are both lengths.

Similarly, the 2-periodic orbit with 3 reflections has length $6R$ while
the periodic orbit with 3 reflections at 3 different points has length
$4\sqrt{2}R<6R$.\\

\noindent
{\bf 4.2 Multiply connected domains}\\

Once again, we consider first rays from $\x$ to $\y$ that are
reflected only once in the boundary. For every connected component
of $\p \Omega $, denoted $\p \Omega _{i}\ (i=1,\dots ,I)$, a point
$\x$  in $\Omega$ is regular with respect to $\p \Omega _{i}$ if
there is only one minimal eikonal $S_{i,1}(\x,\x)>0$ with one
reflection at $\p \Omega _{i}$. We denote by $\Gamma_{i}$ the
locus of the irregular points of $\Omega$ with respect to $\p
\Omega _{i}$.

As above, we define in $\Omega-\Gamma_i$ the minimal eikonal with
one reflection in $\p \Omega _{i}$ such that
$S_{i,2}(\x,\x)>S_{i,1} (\x,\x)$. We construct an approximation
\[G(\y,\x,t)\sim
G_0(\y,\x,t)+ \sum_{k=1}^2\sum_{i=1}^IG_{i,k}(\y, \x,t)\] where
$G_{i,k}(\y,\x,t)$ are ray solutions with eikonals
$S_{i,k}(\y,\x)$ and $Z_{i,k}(\y,\x)$ chosen so as to minimize the
boundary values of the sum at the boundary points of rays
orthogonal to the boundary, as above. The trace of the double sum
is calculated by introducing the change of variables
$2r_{i,k}(\x)=S_{i,k}(\x, \x)$ and arclength $s_i(\x)$ in
$\p\Omega_i$, as above. The Laplace evaluation of the integrals
produces exponents that are the 2-periodic orbits in $\Omega$.


Eikonals with two or more reflections contribute exponents that are
lengths of extremal closed orbits with any number of reflections in the boundary,
as in the case of simply connected domains. Thus the exponents $\delta_n$ in
(\ref{bao}) consist of half the lengths of 2-periodic orbits in $\Omega$ and their multiples, and extremal
lengths of closed periodic orbits with any number of reflections in the boundary and their multiples.\\

\noindent {\large {\bf 5. Discussion}}\newline
\renewcommand{\theequation}{5.\arabic{equation}}
\setcounter{equation}{0} %
\renewcommand{\thefigure}{5.\arabic{figure}} \setcounter{figure}{0}

First, we compare our result to that conjectured in \cite{Berry}.
The leading exponent in the remainder of the short time expansion
of $P(t)$ in an asymptotic power series and that in the large $s$
expansion of its Laplace transform are related by the well known
formula \cite{SA}
\[{\cal L}\left[
\mbox{erfc}\left(\frac{k}{2\sqrt{t}}\right)\right](s)=\frac{1}{s}
e^{-k\sqrt{s}}, \]
where
\[\mbox{erfc}(z)\sim\frac{e^{-z^2}}{\sqrt{\pi}z}\quad\mbox{for
$z\gg1$}.\]
Thus the
exponential rate of blowup of the Laplace transform on a Stokes line is
twice the square root of the exponential decay rate of the remainder
term in
the expansion of the trace. It follows from our result that the
exponential
blowup rate of the Laplace transform is twice the square root of the
exponential rate of decay in the time domain.

We illustrate our expansion for a disk, whose boundary has only
one connected component and a single critical point. We consider
points $\x=\left( x_{1},y_{1}\right) $ and $\y=\left(
x_{2},y_{2}\right) $ inside a circle of radius $R$ centered at the
origin. The leading order eikonal is
\[
S_{0}\left( \y,\x\right) =\left| \x-\y\right| .
\]%
When both $\x$ and $\y$ are on the
$x$-axis, we have $y_{1}=y_{2}=0$ and $%
S_{0}\left( \y,\x\right) =\left|
x_{1}-x_{2}\right| $. Denoting $%
\x_{1}=\left( x_{1},0\right) $ and
$\x_{2}=\left( x_{2},0\right) $%
, we see that the values of the eikonal on the $x$-axis are
$S_{0}\left( \x_{1},\x_{2}\right) =\left|
x_{1}-x_{2}\right| $. We assume that $%
x_{1}>0 $. The boundary values of the eikonal are%
\[
S_{0}\left( \x_{1},\x_{2}\right)
=R-x_{1}\quad \mbox{at \ }\x%
_{2}=(R,0)
\]%
and
\[
S_{0}\left( \x_{1},\x_{2}\right)
=R+x_{1}\quad \mbox{at \ }\x%
_{2}=(-R,0).
\]%
Thus the leading order ray approximation to Green's function
$G\left( \y,\x,t\right)$,
\[
G_{0}\left( \y,\x,t\right) =\frac{1}{4\pi t}e^{-\displaystyle
S_{0}^{2}\left( \y,\x\right) /4t},
\]%
misses the boundary conditions when $\x$ and
$\y$ are on the $x$%
-axis, giving
\[
G_{0}\left( \x_{1},\x_{2},t\right)
=\frac{1}{4\pi t}%
e^{-\left( R-x_{1}\right) ^{2}/4t}\quad \mbox{at }\
\x_{2}=(R,0) \]%
and%
\begin{equation}
G_{0}\left( \x_{1},\x_{2},t\right)
=\frac{1}{4\pi t}%
e^{-\left( R+x_{1}\right) ^{2}/4t}\quad \mbox{at $\x_{2}=(-R,0)$}.
\label{G0error}
\end{equation}

The next eikonal, denoted $S_{1}\left( \y,\x\right) $, is given on
the $x$-axis by $S_{1}\left( \x_{1},\x_{2}\right)
=2R-x_{1}-x_{2},$
and its boundary values are%
\[
S_{1}\left( \x_{1},\x_{2}\right)
=R-x_{1}\quad \mbox{at \ }\x%
_{2}=(R,0)
\]%
and
\[
S_{1}\left( \x_{1},\x_{2}\right)
=3R-x_{1}\quad \mbox{at \ }\x%
_{2}=(-R,0).
\]%
Thus the approximation of Green's function $G\left(
\y,\x,t\right),$
\[
G\left( \y,\x,t\right) \sim G_{0}\left( \y,\x,t\right)
-G_{1}\left( \y,\x,t\right) ,
\]%
corresponding to the ray solutions $G_{0}\left( \y,\x,t\right) $
and%
\[
G_{1}\left( \x,\y,t\right) =Z_{1}\left( \x,\y,t\right)
e^{-\displaystyle S_{1}^{2}\left( \y,\x\right) /4t},
\]%
will satisfy the boundary condition at $\left( x_{1},R\right) $ if $%
Z_{1}\left( \y,\x,t\right) $ is chosen
so that%
\[
Z_{1}\left( \x_{1},\x_{2}{\bf
,}t\right) =\frac{1}{4\pi t}\quad %
\mbox{at \ }\x_{2}=(R,0).
\]%
However, this approximation does not satisfy the boundary
condition at $\x_{2}=(-R,0)$. The error in the boundary values at
$\x_{2}=(-R,0)$ is
\[
G_{0}\left( \x_{1},\x_{2}{\bf
,}t\right) -G_{1}\left( \x_{1},%
\x_{2}{\bf ,}t\right) =\frac{1}{4\pi t}e^{-\left( R+x_{1}\right)
^{2}/4t}-Z_{1}\left( \x_{1},\x_{2}{\bf ,}t\right) e^{-4\left(
R-x_{1}\right) ^{2}/4t},\quad \mbox{at\ }
\x_{2}=(-R,0) \]%
and is of the same order of magnitude as that of the leading order
approximation (\ref{G0error}). To make up for the missed boundary
condition
the further approximation%
\begin{equation}
G\left( \y,\x,t\right) \sim G_{0}\left( \y,\x,t\right)
-G_{1}\left( \y,\x,t\right) -G_{2}\left( \y,\x,t\right) \label{G4}
\end{equation}%
can be used, with%
\[
G_{2}\left( \y,\x,t\right) =Z_{2}\left( \y,\x,t\right)
e^{-\displaystyle s_{1}^{2}\left( \y,\x\right) /4t},
\]%
where on the $x$-axis
\[
s_{1}\left( \x_{1},\x_{2}\right) =2R+x_{1}+x_{2}
\]%
and%
\[
Z_{2}\left( \x_{1},\x_{2}{\bf
,}t\right) =\frac{1}{4\pi t}\quad %
\mbox{at \ }\x_{2}=(-R,0).
\]%
This eikonal corresponds to rays with two reflections in the boundary.
The
approximation (\ref{G4}) decreases the error in the boundary condition
at $%
\x_{2}=(-R,0)$ to
\[-Z_{1}\left( \x_{1},\x_{2}{\bf
,}t\right) e^{-4\left( R-x_{1}\right) ^{2}/4t},\] but misses the
boundary condition at $\x_{2}=(R,0)$ with error
\[
G_{0}\left( \x_{1},\x_{2}{\bf
,}t\right) -G_{1}\left( \x_{1},%
\x_{2}{\bf ,}t\right) -G_{2}\left(
\x_{1},\x_{2}{\bf ,}%
t\right) =-Z_{2}\left( \x_{1},\x_{2}{\bf ,}t\right) e^{-\left(
3R+x_{1}\right) ^{2}/4t}\quad \mbox{at \ }\x_{2}=(R,0).
\]%
This process gives successive approximations to Green's function with
errors
that decrease at transcendental rather than algebraic rates.

The approximation to the trace produced by $G_{0}\left(
\y,\x,t\right)$ is the first algebraic term in the expansion
(\ref{sum}). The contributions of the terms $-G_{1}\left(
\y,\x,t\right)$ and $-G_{2}\left( \y,\x,t\right)$ in the
approximation (\ref{G4}) of terms that are $O\left(\sqrt{t}
e^{-R^{2}/t}\right)$ are identical, but with opposite signs and
thus they cancel each other. The second term contributes a
negative term that is $O\left(\sqrt{ t}e^{-4R^{2}/t}\right) $. The
term $O\left( \sqrt{ t}e^{-R^{2}/t}\right) $ for small $t$
corresponds to $O\left(\displaystyle
\frac{1}{{s}}e^{-2R\sqrt{s}}\right) $ for large positive $s$ in
the Laplace plane. The number $2R$ is the length of the periodic
orbit of a billiard ball bouncing inside a circle with the center
removed, that is, inside the domain $\Omega -\Gamma $, where the
set of
critical points $\Gamma $ consists of the center. Similarly, the term $%
O\left(\sqrt{ t}e^{-4R^{2}/t}\right) $ for small $t$
corresponds to $O\left( \displaystyle \frac{1}{{s}}e^{-4R\sqrt{s}%
}\right) $ for large positive $s$ in the Laplace plane. The number $4R$
is the length of the minimal periodic orbit of a billiard ball bouncing
inside a disk. We conclude that the conjecture of \cite{Berry} should be
supplemented with the orbit of length $2R$.

If $\Omega $ is an annulus between two concentric circles, of radii $a$ and $b$,
respectively, ($a>b$), the two connected components of the boundary are the two circles and there
are no critical points in the domain relative to either one of them. In this case $\delta_{1}=(a-b)$.

If $\Omega $ is the ellipse
\[\frac{x^2}{a^2}+\frac{y^2}{b^2}<1\]
with $a>b$, the locus of critical points relative to the boundary is the segment
$$\Gamma=\displaystyle\left[-\frac{a^2-b^2}{a},\frac{a^2-b^2}{a}\right]$$ on the $x$-axis. The segment $\Gamma$ is the short diagonal of the evolute of the ellipse (the asteroid
$\left(ax\right)^{2/3}+\left(by\right)^{2/3}=\left(a^2-b^2\right)^{2/3}$).
For the ellipse there are exponents in eq.(\ref{bao}) which are $\delta_1=2b$ and its multiples and $\delta_2=2a$ and
its multiples, as well as extremal periodic orbits with any number of reflections in the boundary.

Finally, we observe that if the boundary is reflecting (i.e., a
homogeneous Neumann boundary condition), the exponential decay rate of the
transcendental terms in the expansion of the trace is the same as in the
case of absorbing boundary (homogeneous Dirichlet boundary condition).
In this case the second term in the expansion (\ref{bao}) changes sign.

Obviously, rays that are reflected from the boundary more than once also
give rise to ray solutions. The number of ray solutions needed in the
expansion (\ref{Raysolution1}) is determined by the required degree of
asymptotic approximation of the boundary conditions. If only a finite
sum of ray solutions satisfies the boundary conditions, the sum
(\ref{Raysolution1}) is finite. Otherwise, additional ray solutions improve
the degree of approximation of the boundary conditions, as described in the
one-dimensional ray expansion in Section 2.

Finally, the asymptotic convergence of the ray expansion follows
from the maximum principle for the heat equation in a
straightforward manner.

\end{document}